\documentclass[conference]{IEEEtran}
\IEEEoverridecommandlockouts
\usepackage{cite}
\usepackage{amsmath,amssymb,amsfonts}
\usepackage{algorithmic}
\usepackage{graphicx}
\usepackage{textcomp}
\usepackage{xcolor}
\def\BibTeX{{\rm B\kern-.05em{\sc i\kern-.025em b}\kern-.08em
		T\kern-.1667em\lower.7ex\hbox{E}\kern-.125emX}}

\usepackage{adjustbox}
\usepackage{amsmath}
\usepackage{amssymb}
\usepackage{amsthm}
\usepackage{graphicx}
\usepackage{pstricks}
\usepackage{pst-grad} 
\usepackage{pst-plot} 
\usepackage{multirow}
\usepackage{epsfig}
\usepackage{url}
\usepackage{lscape}
\usepackage{float}
\usepackage{cases}
\usepackage{booktabs}
\usepackage{verbatim}
\usepackage{latexsym}
\usepackage{mathrsfs}
\usepackage{algorithm}
\usepackage{algorithmic}
\usepackage{syntonly}
\usepackage{enumitem}
\usepackage{moresize}
\usepackage[font=small]{subfig}
\usepackage[font=footnotesize, labelfont=bf, justification=centering]{caption}
\usepackage{bm}
\usepackage{listings}
\usepackage{bbm}
\usepackage{color}
\usepackage{lscape}

\usepackage{hyperref} 

\begin{document}
	
	\title{Classification Beats Regression: Counting of Cells from Greyscale Microscopic Images based on Annotation-free Training Samples}
	

	 \author{\IEEEauthorblockN{1\textsuperscript{st} Xin Ding}
	 \IEEEauthorblockA{\textit{Department of Statistics} \\
	 \textit{University of British Columbia}\\
	 Vancouver, CA \\
	 xin.ding@stat.ubc.ca}
	 \and
	 \IEEEauthorblockN{1\textsuperscript{st} Qiong Zhang}
	 \IEEEauthorblockA{\textit{Department of Statistics} \\
	 \textit{University of British Columbia}\\
	 Vancouver, CA \\
	 qiong.zhang@stat.ubc.ca}
	 \and
	 \IEEEauthorblockN{2\textsuperscript{nd} William J. Welch}
	 \IEEEauthorblockA{\textit{Department of Statistics} \\
	 \textit{University of British Columbia}\\
	 Vancouver, CA \\
	 will@stat.ubc.ca}
	 }
	
	\maketitle
	\thispagestyle{plain} 
	\pagestyle{plain} 
	
	\begin{abstract}
		
		Modern methods often formulate the counting of cells from microscopic images as a regression problem and more or less rely on expensive, manually annotated training images (e.g., dot annotations indicating the centroids of cells or segmentation masks identifying the contours of cells). This work proposes a supervised learning framework based on classification-oriented convolutional neural networks (CNNs) to count cells from greyscale microscopic images without using annotated training images. In this framework, we formulate the cell counting task as an image classification problem, where the cell counts are taken as class labels. This formulation has its limitation when some cell counts in the test stage do not appear in the training data. Moreover, the ordinal relation among cell counts is not utilized. To deal with these limitations, we propose a simple but effective data augmentation (DA) method to synthesize images for the unseen cell counts. We also introduce an ensemble method, which can not only moderate the influence of unseen cell counts but also utilize the ordinal information to improve the prediction accuracy. This framework outperforms many modern cell counting methods and won the data analysis competition (\textit{Case Study 1: Counting Cells From Microscopic Images}\footnote{\url{https://ssc.ca/en/case-study/case-study-1-counting-cells-microscopic-images}}) of the 47th Annual Meeting of the Statistical Society of Canada (SSC). Our code is available at \url{https://github.com/UBCDingXin/CellCount_TinyBBBC005}. 
		
	\end{abstract}
	
	
	\section{Introduction}\label{sec:intro}
	
	Modern methods \cite{xie2018microscopy, liu2019novel, hernandez2018using, liu2019automated, xue2016cell, xue2018cell} often formulate the counting of cells from microscopic images as a regression problem, where the cell count is regressed on some image-related covariates. Most of these regression-based methods \cite{xie2018microscopy, liu2019novel, hernandez2018using, liu2019automated} can be generalized as a two-step procedure: extract some high-level features as the covariates from a microscopic image, and then regress the cell count on these covariates. At the first step, \cite{xie2018microscopy} and \cite{liu2019novel} extract a dot density map \cite{lempitsky2010learning} from an image via a U-net \cite{ronneberger2015u}, while \cite{hernandez2018using} generates a segmentation mask for cells in an image via a Feature Pyramid Network (FPN) \cite{lin2017feature}. At the second step, \cite{xie2018microscopy} directly sums up the dot density map to fit the cell count; \cite{liu2019novel} and \cite{hernandez2018using} regress the cell count on the dot density map and the segmentation mask, respectively,  by a VGG-19 net \cite{simonyan2014very}. The VGG-19 net \cite{simonyan2014very} is also utilized by \cite{liu2019automated} to regress the cell count on an integration of the dot density map and the segmentation mask. These two-step methods require manual annotations (i.e., dot annotations indicating the centroids of cells or segmentation masks identifying the contours of cells) on the training images. Unfortunately, it is usually difficult and expensive to obtain such annotations in practice. To avoid these expensive manual annotations, \cite{xue2016cell} and \cite{xue2018cell} propose to directly regress the cell count on the microscopic image via residual neural networks (ResNets) \cite{he2016deep}. However, all these regression-based approaches have the drawback of rarely making a precise prediction, i.e., the predicted cell counts often deviate from the ground truth even for some simple test images. 
	
	Instead of the regression-based formulation, we can formulate the cell counting problem as one of classification, where the cell counts are taken as class labels. In this formulation, we predict the cell count (i.e., class label) directly from the greyscale microscopic images by using classification-oriented CNNs. This formulation has two advantages. First, it does not require any manual annotation (e.g., dot annotation or segmentation mask) on the training images that is expensive to obtain in practice. Second, as long as a test image is correctly classified, the prediction error on it is exactly zero. However, besides these two advantages, there are also two limitations of this formulation which explain why the modern cell counting methods are all regression-oriented. First, if a test image has a cell count not seen in the training set, then a prediction error is inevitable. Second, since the ordinal information in the cell counts is not utilized by the classification-oriented CNNs (e.g., counts 10 and 11 are taken as far apart as counts 10 and 100), even when the classification loss (e.g., cross entropy) on a test image is small, the predicted cell count may still be far from the ground truth. 	
	
	We therefore introduce a novel framework to count cells from greyscale microscopic images to tackle the drawbacks of existing methods and the limitations of the classification-based formulation. Our contributions can be summarized as follows:
	\begin{itemize}
		\item We propose in Section \ref{sec:method} a novel framework to count cells from greyscale microscopic images in which the cell counting problem is formulated as an image classification. This framework consists of a simple but effective data augmentation (DA) to synthesize images for unseen cell counts. An ensemble scheme (i.e., combine a classification-oriented CNN with a regression-oriented CNN) is also included in this framework to not only deal with the unseen cell counts but also utilize the ordinal information to avoid large prediction errors.
		\item In Section \ref{sec:experiment}, we introduce the Tiny-BBBC005 dataset (used by  the data analysis competition of the 47th Annual Meeting of the SSC), on which the proposed framework outperforms several modern cell counting methods. The simple data augmentation and the ensemble scheme can effectively alleviate the two limitations of the classification-based formulation.
	\end{itemize}

	\section{Related Works}\label{sec:related_works}
	In this section, we review some modern cell counting methods, which are empirically compared with our proposed framework in Section \ref{sec:experiment}.
	
	{\setlength{\parindent}{0cm}\textbf{DRDCNN \cite{liu2019novel}:}} DRDCNN is a two-step method. A U-Net is first trained to extract the dot density map with minimum Mean Square Error (MSE) from a given microscopic image. A VGG-19 is then attached to the trained U-Net to predict the cell count based on the extracted dot density map and trained to minimize the MSE loss.

	{\setlength{\parindent}{0cm}\textbf{FPNCNN \cite{hernandez2018using}:}} FPNCNN is another two-step method which first uses a FPN to generate the segmentation mask for cells in a given microscopic image, and then regresses the cell count on this mask via a VGG-19. The FPN is trained to minimize the sum of the aleatoric loss \cite{kendall2017uncertainties} and the total-variational (TV) loss~\cite{chambolle2004algorithm}. Similar to DRDCNN, the VGG-19 is trained to minimize the MSE loss.

	{\setlength{\parindent}{0cm}\textbf{ERDCNN \cite{liu2019automated}:}} ERDCNN is a combination of DRDCNN and FPNCNN, where the cell count is regressed on an integration of the dot density maps and the segmentation masks. Different from FPNCNN, the segmentation masks are generated by a U-Net instead of a FPN. ERDCNN also uses a VGG-19 to predict cell counts with MSE loss.

	{\setlength{\parindent}{0cm}\textbf{Regression-oriented ResNets \cite{xue2016cell,xue2018cell}:}} \cite{xue2016cell} and \cite{xue2018cell} propose to directly predict the cell count from a microscopic image by using regression-oriented ResNets, which are trained to minimize the MSE loss.

	\section{Method}\label{sec:method}	
	\subsection{Overview}\label{sec:overview}
\begin{figure}[ht]
		\centering
		\subfloat[][Training stage. Please see Section \ref{sec:counting_by_class_cnns} and \ref{sec:ensemble} for the definitions of ResNet-XX (CE), ResNet-XX (MSE), and LQReg.]{
			\includegraphics[width=0.95\linewidth]{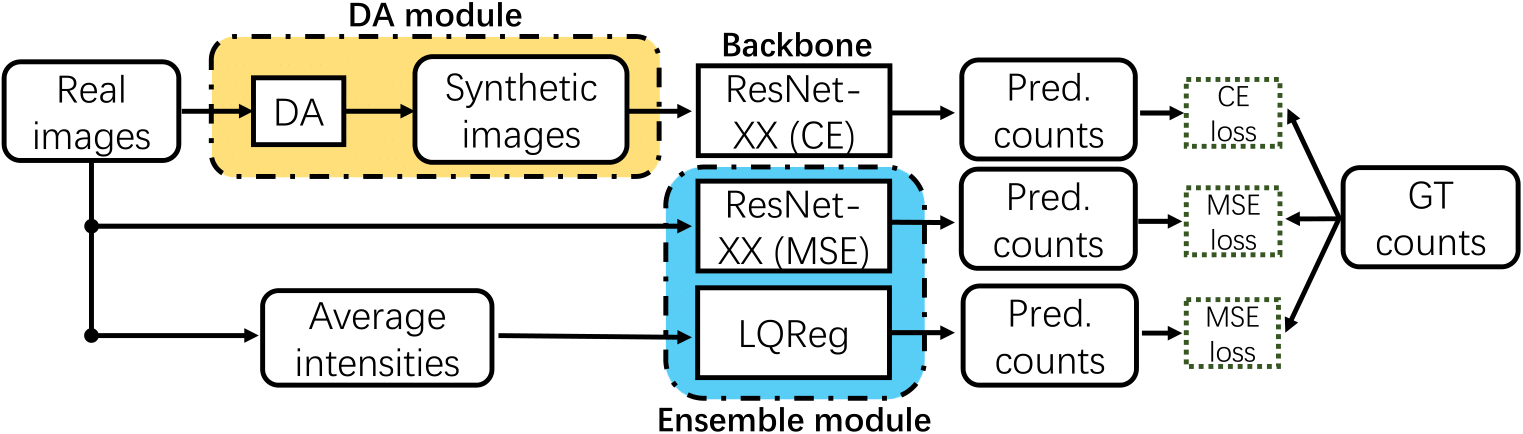}
			\label{fig:workflow_train}}\\
		\subfloat[][Testing stage. If the ensemble module is disabled, then the final prediction is Pred. count 1.]{
			\includegraphics[width=0.95\linewidth]{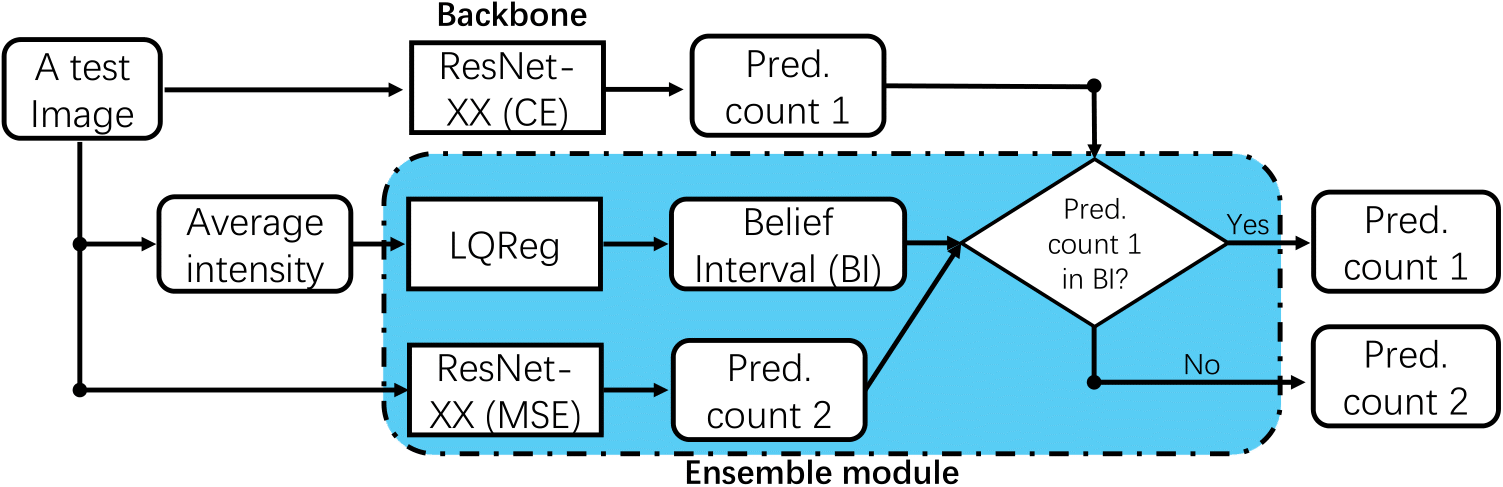}
			\label{fig:workflow_test}}
		\caption{The workflows for our proposed framework in the training and testing stages. The data augmentation (DA) module is enabled if we know which cell counts are missing. The ensemble module is enabled if ResNet-XX (CE) does not work well due to either missing cell counts (known or unknown) or insufficient training samples.}
		\label{fig:workflow}
	\end{figure}

	In this section, we propose a novel framework to count cells from greyscale microscopic images. We first formulate the cell counting task as a classification problem, where the cell counts are taken as class labels. Then, we use modern classification CNNs (e.g., ResNets) as the backbone to predict cell counts from greyscale microscopic images. A simple but effective data augmentation and an ensemble scheme are also proposed to 
	deal with the two limitations of the classification CNNs (discussed in Section \ref{sec:intro}). The workflows of our proposed framework in the training and testing stages are visualized, respectively, in Figs.\  \ref{fig:workflow}(a) and \ref{fig:workflow}(b). 
	
	\subsection{Counting Cells by Classification-oriented CNNs}\label{sec:counting_by_class_cnns}
	As the backbone of our framework, we count cells from greyscale microscopic images by classification-oriented ResNets which are trained to minimize the Cross Entropy (CE) loss. These ResNets take the greyscale microscopic image as input and output the cell counts (i.e., class labels). To distinguish between the regression-oriented ResNets in \cite{xue2016cell,xue2018cell} and the classification-oriented ResNets in our framework, we denote all ResNets in this paper as ResNet-XX (YY), where XX and YY, respectively, represent the number of convoluational or linear layers and the loss function. For example, the ResNet-34 (MSE) and ResNet-34 (CE) in Section \ref{sec:experiment} are trained for regression and classification respectively.
	
	{\setlength{\parindent}{0cm}\textbf{Two limitations:}} As we discussed in Section \ref{sec:intro}, there are two limitations of the classification-oriented formulation. First, some cell counts in the test set may be missing in the training set, so prediction errors on these test samples are inevitable. Second, the ordinal information in the cell counts is not utilized, which may consequently lead to large prediction errors on some test images.
	
	\subsection{A Simple But Effective Data Augmentation}\label{sec:data_augmentation}
	
\begin{figure}[ht]
		\centering
		\subfloat[][5 Cells (real)]{
			\includegraphics[width=2cm, height=2cm]{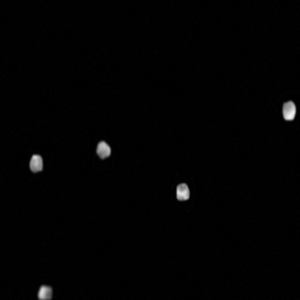}
			\label{fig:DA_img_1}}
		\subfloat[][10 Cells (real)]{
			\includegraphics[width=2cm, height=2cm]{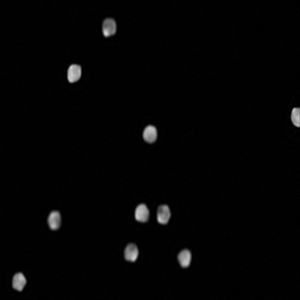}
			\label{fig:DA_img_2}}
		\subfloat[][15 Cells (synthetic)]{
			\includegraphics[width=2cm, height=2cm]{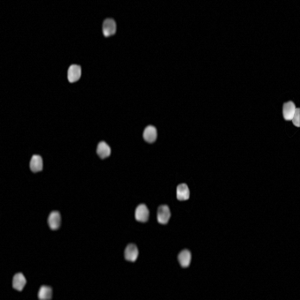}
			\label{fig:DA_img_3}}
		\caption{Creating a synthetic greyscale microscopic image with 15 cells by overlaying two real Tiny-BBBC005 images with 5 and 10 cells respectively.}
		\label{fig:example_DA}
	\end{figure}

	If we know which cell counts in the test set do not appear in the training data, we propose a simple but effective data augmentation (DA) to synthesize greyscale microscopic images for these missing cell counts. 
	Given distinct images $X^{(1)}, \dots, X^{(n)}$ and their cell counts $y^{(1)}, \cdots, y^{(n)}$, a synthetic image $Z$ with cell count $\sum_{k=1}^ny^{(k)}$ can be created via 
	\begin{equation}
		\label{eq:DA}
		Z_{ij} = \max\{X^{(1)}_{ij}, X^{(2)}_{ij}, \cdots, X^{(n)}_{ij}\},
	\end{equation}
	where $Z_{ij}$ is the $(i,j)$-th pixel of $Z$. We propose to use the max operation in Eq.\eqref{eq:DA} rather than a convex combination to overlay multiple images since the synthetic images obtained via the convex combination usually lack contrast. Additionally, the $\max$ operation guarantees the output pixel values do not exceed the bit depth of real images. An example of creating a synthetic image with 15 cells by overlaying two real images from Tiny-BBBC005 with 5 and 10 cells respectively is shown in Fig.~\ref{fig:example_DA} and the synthetic image (i.e., Fig.~\ref{fig:example_DA}(c)) with 15 cells looks very realistic.
	
	In Fig.~\ref{fig:example_DA}, an unseen cell count 15 is created based on a combination of two existing images with cell counts 5 and 10 (known as ``basis cell counts"). Such a synthesis can be written as a formula, i.e., $15 = 5\times(1) + 10\times(1)$, 
	where the first number in a product is a basis cell count and the associated number in parentheses is the number of distinct real images with that count. In addition to which operation should be used for overlaying multiple images, another question is which formula should be selected to create an unseen cell count? Often there will be more than one possible formula. For example, to create 15-cell images, we can use the following formulae: $15 = 1\times(15)$, $15 = 1\times(5) + 10\times(1)$, $15 = 5\times(1) + 10\times(1)$, $15 = 1\times(10) + 5\times(1)$, etc.
	If there is enough computation budget, we suggest taking into account as many potential formulae as possible to increase the variety of synthetic images. For a given unseen cell count, we can create a pool of potential formulae. One synthetic image with this unseen cell count is synthesized at a time by randomly choosing one formula from the pool.

	\subsection{Ensembling Classification and Regression Methods}\label{sec:ensemble}
	
	To further improve the prediction accuracy of our framework, we also propose an ensemble scheme which combines the high precision of classification-oriented ResNets on seen cell counts with the high stability of regression-oriented ResNets on both seen and unseen cell counts. To be specific, for a test image, we first predict the cell count by ResNet-XX (CE); however, if the predicted cell count from ResNet-XX (CE) is outside a certain interval (termed the \textit{belief interval}), then we use ResNet-XX (MSE) to make the prediction. 
	
\begin{figure}[!htbp]
		\centering
		\subfloat[][Linear regression (LReg) lines for nuclei stained and blur level 1 images]{
			\label{fig:LReg_bound}
			\includegraphics[width=0.47\linewidth]{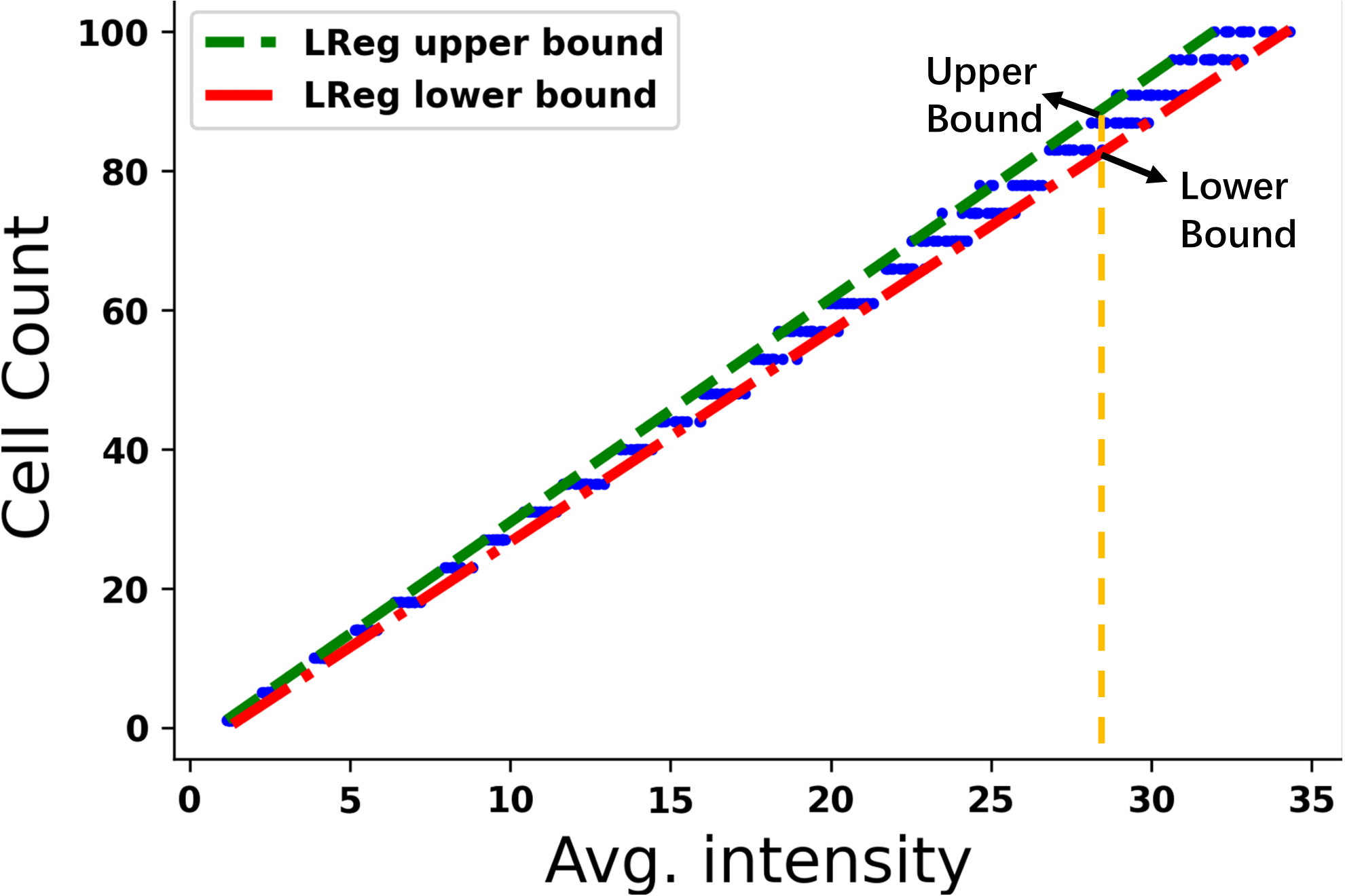}}
		\subfloat[][Quadratic regression (QReg) lines for body stained and blur level 1 images]{
			\label{fig:QReg_bound}
			\includegraphics[width=0.47\linewidth]{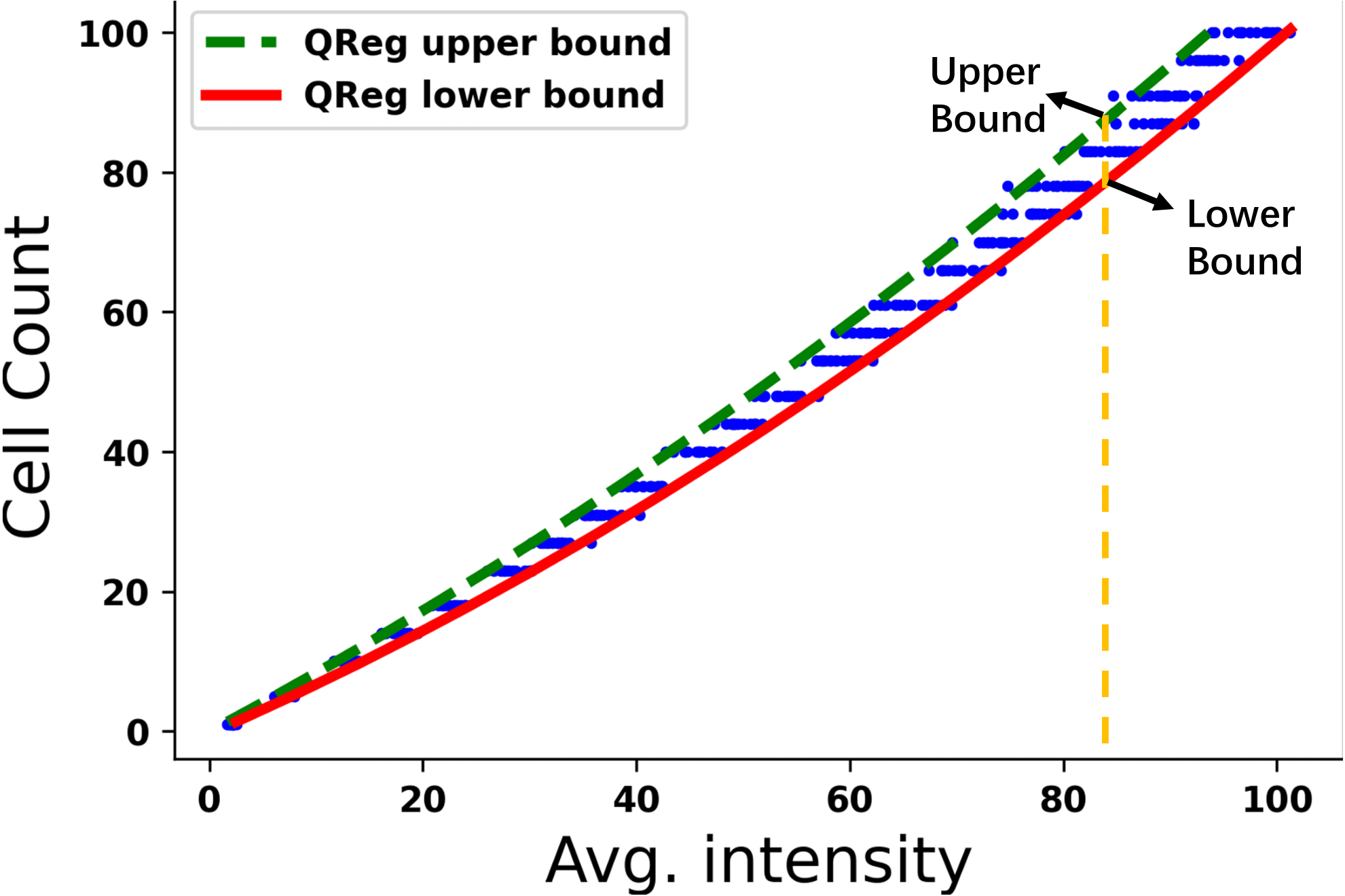}}
		\caption{Scatter plots (blue dots) of the nuclei/body stained and blur level 1 Tiny-BBBC005 images with linear/quadratic regression lines. The green regression line is fitted to data points close to the $y$-axis while the red regression line is fitted to data points close to the $x$-axis. The green and red regression lines provide a belief interval for the ground truth cell count of any given image.}
		\label{fig:ensemble_LQReg_lines}
	\end{figure}

	We use Tiny-BBBC005 as an example to show how to build the belief interval for the ensemble, which may be generalized to other greyscale microscopic image datasets. Specifically, from Fig.~\ref{fig:example_DA}, we can see that greyscale images with larger cell counts usually have higher average intensity (i.e., the average pixel value of an image). Fig.~\ref{fig:ensemble_LQReg_lines}(a) shows a scatter plot of the nuclei stained images and blur level 1 in Tiny-BBBC005, with average intensity on the $x$-axis and cell count on the $y$-axis (the different types of stain and the blur levels are introduced in Section \ref{sec:dataset}). Fig.~\ref{fig:ensemble_LQReg_lines}(b) shows the analogous plot for body stained images with blur level 1.  We can see that Fig.~\ref{fig:ensemble_LQReg_lines}(a) implies a linear relation between average intensity and cell count, while Fig.~\ref{fig:ensemble_LQReg_lines}(b) implies a quadratic relation. Moreover, in both Fig.~\ref{fig:ensemble_LQReg_lines}(a) and Fig.~\ref{fig:ensemble_LQReg_lines}(b), we can see there is a smallest average intensity and a largest average intensity for each cell count. Based on these findings, for each of the three nuclei stained image groups in the training set of Tiny-BBBC005, we fit two linear regression models, i.e., the green and red regression lines in Fig.~\ref{fig:ensemble_LQReg_lines}(a). The green regression line is fitted on images with the smallest average intensities (i.e., data points close to the $y$-axis) while the red regression line is fitted on images with the largest average intensities (i.e., data points close to the $x$-axis). Similarly, for each of the three body stained image groups, we fit two quadratic regression models. After fitting these linear/quadratic regression models, given a test image along with its stain type and blur level, we first compute the average intensity of this image and plug it into the linear/quadratic regression models to get a predicted upper bound and a predicted lower bound  (shown in Fig.~\ref{fig:ensemble_LQReg_lines}) for the ground truth cell count of this image. These lower and upper bounds form the belief intervals in the ensemble.
	
	The proposed ensemble scheme has two benefits. First, it moderates the influence of unseen cell counts on ResNet-XX (CE), even when we do not know which counts are missing. Second, the belief interval implicitly utilizes the ordinal information in the cell counts and ensures that the ResNet-XX (CE) does not make ``big" mistakes.
	
	\section{Experiment}\label{sec:experiment}
	We first introduce the Tiny-BBBC005 dataset, some implementation details of our experiments, and two evaluation metrics. Then, we compare our proposed framework with several modern methods (summarized in Section \ref{sec:related_works}) under four experimental scenarios: (1) all cell counts in the test set appear in the training set; (2) some randomly selected cell counts along with their images from the training set are deleted; (3) five consecutive cell counts along with their images from the training set are deleted; and (4) half of the training images, chosen at random, are deleted. 
	
	\subsection{Dataset: Tiny-BBBC005}\label{sec:dataset}
	Tiny-BBBC005, used in Case Study 1 of the 47th Annual Meeting of the SSC, is a subset of the Broad Bioimage Benchmark Collection (BBBC005)~\cite{ljosa2012annotated} and consists of $3600$ simulated greyscale microscopic images~\cite{lehmussola2007computational,lehmussola2008synthetic} of size $696\times 520$ for $600$ plates of cells. There are $24$ distinct cell counts (i.e., 1,   5,  10,  14,  18,  23,  27,  31,  35,  40,  44,  48,  53, 57,  61,  66,  70,  74,  78,  83,  87,  91,  96, 100) that are evenly distributed across $1$ to $100$. For each image, a ground truth segmentation mask that identify the contours of cells is also provided, but Tiny-BBBC005 does not have ground truth dot density maps.
	
\begin{figure}[ht]
		\centering
		\subfloat[][Nuclei stain (in-focus; blur level 1)]{
			\includegraphics[width=2cm, height=2cm]{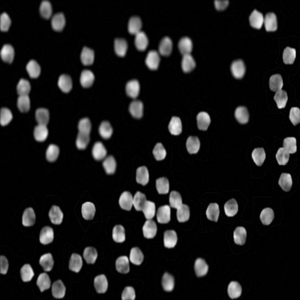}
			\label{fig:samp_img_1}}
		\subfloat[][Body stain (in-focus; blur level 1)]{
			\includegraphics[width=2cm, height=2cm]{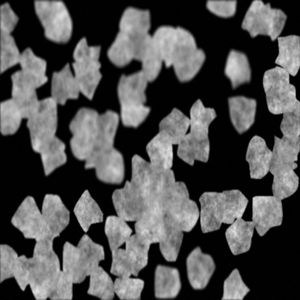}
			\label{fig:samp_img_2}}
		\subfloat[][Body stain (out-of-focus; blur level 23)]{
			\includegraphics[width=2cm, height=2cm]{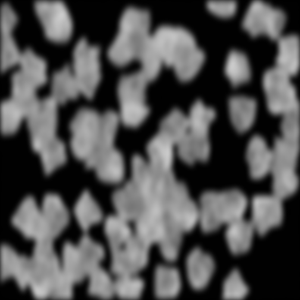}
			\label{fig:samp_img_3}}
		\subfloat[][Body stain (out-of-focus; blur level 48)]{
			\includegraphics[width=2cm, height=2cm]{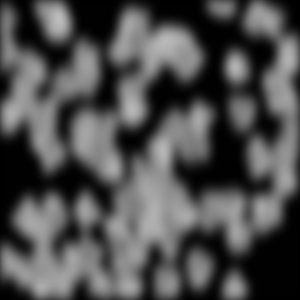}
			\label{fig:samp_img_4}}
		\caption{Four simulated microscopic images from Tiny-BBBC005 for the same plate of 78 cells with different stain type and blur level. Figure~\ref{fig:samp_img_1} and~\ref{fig:samp_img_2} are in-focus images with blur level 1. Figs.~\ref{fig:samp_img_3} and~\ref{fig:samp_img_4} are out-of-focus images with blur level 23 and 48 respectively.}
		\label{fig:example_figures}
	\end{figure}

	To separate cells from the background and facilitate counting, cells are usually dyed in practice. There are two common methods: nuclei stain and cell body stain. In Tiny-BBBC005, all the cells of each plate are dyed in both ways. For example, Fig.~\ref{fig:samp_img_1} and Fig.~\ref{fig:samp_img_2} are respectively the nuclei stained and body stained image of the same plate with $78$ cells. For the nuclei stained images, the cells are better separated, whereas cell overlapping is more severe for body stained images. There are also 3 levels (i.e., 1, 23 and 48) of focus blur for an image. Focus blur was simulated by applying Gaussian filters to in-focus images. For example, Fig.~\ref{fig:samp_img_3} and Fig.~\ref{fig:samp_img_4} are the out-of-focus images simulated from the in-focus image in Fig.~\ref{fig:samp_img_2}. Therefore, each plate of cells has 6 images which correspond to the combinations of 3 blur levels and 2 stain types. For each combination of blur level and stain, the 600 images are divided equally between the 24 distinct cell counts, i.e., 25 images per distinct cell count. Note that the blur level and stain type of each image in Tiny-BBBC005 are known to us. 
	
	The 600 images for each combination of stain type and blur level are randomly split into two parts; 400 images form a training set and the remaining 200 comprise the test set. Overall, then, the training set has 2,400 images while the test set has 1,200 images, and these two sets contain the same 24 distinct cell counts.

	\subsection{Implementation Details}\label{sec:implementation}
	We resize all Tiny-BBBC005 images to $256\times 256$ with the Lanczos filter \cite{duchon1979lanczos} by \texttt{Pillow} \cite{clark2015pillow}. Since Tiny-BBBC005 does not provide the ground truth dot density maps, the U-Net for density extraction in DRDCNN and ERDCNN is pre-trained on the VGG dataset \cite{lempitsky2010learning}. The formulae for the DA scheme in Experiment 2 and 3 are shown in Table \ref{tab:EXP2_formulae} and \ref{tab:EXP3_formulae} in the supplementary material. 
	
	\subsection{Evaluation Metrics}\label{sec:metrics}
	In the following experiments, all candidate methods are evaluated on the test set of Tiny-BBBC005 by Root Mean Square Error (RMSE), i.e., 
	\begin{equation}
		\label{eq:evaluation_rmse}
		\text{RMSE}=\sqrt{\frac{1}{1200}\sum_{i=1}^{1200}(c_i-\hat{c}_i)^2},
	\end{equation}
	and Mean Absolute Error (MAE), i.e.,
	\begin{equation}
		\label{eq:evaluation_mae}
		\text{MAE}=\frac{1}{1200}\sum_{i=1}^{1200}|c_i-\hat{c}_i|,
	\end{equation}
	where $c_i$ is the true cell count of test image $i$, and $\hat{c}_i$ is the predicted cell count.

	\subsection{Experiment 1: No Unseen Counts}\label{sec:exp1}
	In the first experiment, all images in the training set of Tiny-BBBC005 are available for training. For regression-oriented ResNets \cite{xue2016cell,xue2018cell}, we include ResNet-34 (MSE), ResNet-50 (MSE), and ResNet-101 (MSE) in the comparison. We do not implement the data augmentation and ensemble schemes, because there are no unseen cell counts in this experiment and each distinct cell count has enough samples. The experiment is designed to show that, under these conditions, the classification-oriented ResNets are already good enough to make precise predictions. For our framework, we only include ResNet-34 (CE), ResNet-50 (CE), and ResNet-101 (CE) in the comparison.
	
	The performance of all candidate methods is summarized in Table~\ref{tab:EXP1_main_result}. We can see our proposed classification method substantially outperforms others regardless of the ResNet architecture. Moreover, we find that the ResNet-34 architecture performs better than ResNet-50 and ResNet-101 in both regression and classification. Therefore, in the following experiments, we only use the ResNet-34 architecture.  
	
	\begin{table}[ht]
		\centering
		\caption{Test RMSE and MAE in Experiment 1. In our framework, the DA and ensemble modules are disabled since ResNet-XX (CE) already performs very well and there are no missing cell counts.}
		\begin{adjustbox}{width=0.95\linewidth}
			\begin{tabular}{lp{1cm}lll}
				\toprule
				\textbf{Method} & \textbf{Annot. Type} & \textbf{Pred. Type} & \textbf{RMSE} & \textbf{MAE} \\
				\midrule
				DRDCNN \cite{liu2019novel} & density & Regression & 3.067 & 2.230 \\
				FPNCNN \cite{hernandez2018using} & masks & Regression & 2.817 & 1.727 \\
				ERDCNN \cite{liu2019automated} & density + masks & Regression & 2.877 & 2.114 \\
				\hline
				ResNet-34 (MSE) \cite{xue2016cell} & None  & Regression & 1.378 & 1.001 \\
				ResNet-50 (MSE) \cite{xue2016cell} & None  & Regression & 1.512 & 1.114 \\
				ResNet-101 (MSE) \cite{xue2016cell} & None  & Regression & 2.140 & 1.551 \\
				\hline
				\textbf{Ours} (only ResNet-34 (CE)) & None  & Classification & \textbf{0.400} & \textbf{0.040} \\
				Ours (only ResNet-50 (CE)) & None  & Classification & 0.757 & 0.093 \\
				Ours (only ResNet-101 (CE)) & None  & Classification & 0.841 & 0.098 \\
				\bottomrule
			\end{tabular}%
		\end{adjustbox}
		\label{tab:EXP1_main_result}%
	\end{table}%

	\subsection{Experiment 2: Randomly Missing Counts}\label{sec:exp2}
	
	To test the effectiveness of the proposed data augmentation and the ensemble scheme, we perform a three-round experiment, where, in each round, we randomly remove five cell counts along with their images from the training set. The five deleted cell counts in each round are shown in Table \ref{tab:EXP2_del_class}. The data augmentation and the ensemble scheme in the proposed framework is applied. When conducting the data augmentation, for each missing count and each combination of stain type and blur level, we create 20 synthetic images. Each synthetic image is created based on a randomly selected formula from a pool of $3$, $4$, or $5$ formulae for each missing count. The experimental results averaged over three rounds are shown in Table~\ref{tab:EXP2_main_result}. 
	
	From Table~\ref{tab:EXP2_main_result}, we can see the proposed framework still outperforms the other five methods. In this experiment, an ablation study is conducted to test the effectiveness of each component of our proposed framework and the results are reported in Table \ref{tab:EXP2_ablation_study}. ResNet-34 (CE) does not perform well due to the missing counts; however, both the data augmentation and the ensemble reduce the impact of these missing counts and their combination leads to the largest performance gain.
	
	\begin{table}[h]
		\small
		\centering
		\caption{Randomly delete five counts in each round of Experiment 2}
		\begin{adjustbox}{width=0.5\linewidth}
		\begin{tabular}{llllll}
			\toprule
			& \multicolumn{5}{c}{Deleted Counts} \\
			\midrule
			Round 1 & 14    & 35    & 57    & 66    & 83 \\
			Round 2 & 10    & 31    & 70    & 83    & 91 \\
			Round 3 & 18    & 27    & 44    & 53    & 91 \\
			\bottomrule
		\end{tabular}%
		\end{adjustbox}
		\label{tab:EXP2_del_class}%
	\end{table}%
	
	\begin{table}[!htbp]
		\centering
		\caption{Average test RMSE and MAE over three rounds in Experiment 2 with standard deviations after the ``$\pm$" symbol. All modules in our framework are enabled.}
		\begin{adjustbox}{width=0.8\linewidth}
		\begin{tabular}{lll}
			\toprule
			\textbf{Method} & \textbf{RMSE} & \textbf{MAE} \\
			\midrule
			DRDCNN \cite{liu2019novel} & $4.103\pm 0.213$ & $3.405\pm 0.261$ \\
			FPNCNN \cite{hernandez2018using} & $2.503\pm 0.195$ & $1.902\pm 0.186$ \\
			ERDCNN \cite{liu2019automated} & $3.620\pm 1.300$ & $2.845\pm 1.258$ \\
			ResNet-34 (MSE) \cite{xue2016cell} & $1.707\pm 0.096$ & $1.242\pm 0.101$ \\
			\textbf{Ours} & $\bm{1.103\pm 0.041}$ & $\bm{0.350\pm 0.031}$ \\
			\bottomrule
		\end{tabular}%
		\end{adjustbox}
		\label{tab:EXP2_main_result}%
	\end{table}%
	
	\begin{table}[!htbp]
		\centering
		\caption{An ablation study of our method in Experiment 2.}
		\begin{adjustbox}{width=6.5cm, height=1cm}
		\begin{tabular}{lll}
			\toprule
			\textbf{Method} & \textbf{RMSE} & \textbf{MAE} \\
			\midrule
			ResNet-34 (CE) & $3.045\pm 0.143$ & $1.238\pm 0.056$ \\
			DA & $2.120\pm 0.105$ & $0.615\pm 0.134$ \\
			Ensemble & $1.359\pm 0.032$ & $0.556\pm 0.016$ \\
			DA+Ensemble  & $\bm{1.103\pm 0.041}$ & $\bm{0.350\pm 0.031}$ \\
			\bottomrule
		\end{tabular}%
		\end{adjustbox}
		\label{tab:EXP2_ablation_study}%
	\end{table}%

	\subsection{Experiment 3: Consecutively Missing Counts}\label{sec:exp3}
	This experiment is similar to Experiment 2 except that at each round, we delete five consecutive cell counts along with their images from the training set (shown in Table \ref{tab:EXP3_del_class}). This experiment is also designed to show the effectiveness of the proposed data augmentation and the ensemble scheme. Since the missing cell counts are consecutive, this setup is more challenging than Experiment 2.

	The main comparison in Table \ref{tab:EXP3_main_result} shows that our proposed framework is far better than the existing methods. The ablation study in Table \ref{tab:EXP3_ablation_study} shows that again both the data augmentation and the ensemble scheme are effective.
	
	\begin{table}[!htbp]
		\small
		\centering
		\caption{Delete five neighboring counts in each round of Experiment 3}
		\begin{adjustbox}{width=0.5\linewidth}
		\begin{tabular}{llllll}
			\toprule
			& \multicolumn{5}{c}{Deleted Counts} \\
			\midrule
			Round 1 & 61    & 66    & 70    & 74    & 78 \\
			Round 2 & 70    & 74    & 78    & 83    & 87 \\
			Round 3 & 83    & 87    & 91    & 96    & 100 \\
			\bottomrule
		\end{tabular}%
		\end{adjustbox}
		\label{tab:EXP3_del_class}%
	\end{table}%
	
	\begin{table}[!htbp]
		\centering
		\caption{Average test RMSE and MAE over three rounds in Experiment 3 with standard deviations after the ``$\pm$" symbol. All modules in our framework are enabled.}
		\begin{adjustbox}{width=0.8\linewidth}
		\begin{tabular}{lll}
			\toprule
			\textbf{Methods} & \textbf{RMSE} & \textbf{MAE} \\
			\midrule
			DRDCNN \cite{liu2019novel} & $4.480\pm 0.484$ & $3.509\pm 0.522$ \\
			FPNCNN \cite{hernandez2018using} & $3.606\pm 0.702$ & $2.529\pm 0.325$ \\
			ERDCNN \cite{liu2019automated} & $3.706\pm 0.495$ & $2.785\pm 0.470$ \\
			ResNet-34 (MSE) \cite{xue2016cell} & $2.200\pm 0.399$ & $1.461\pm 0.162$ \\
			\textbf{Ours} & $\bm{1.664\pm 0.275}$ & $\bm{0.621\pm 0.083}$ \\
			\bottomrule
		\end{tabular}%
		\end{adjustbox}
		\label{tab:EXP3_main_result}%
	\end{table}%
	
	\begin{table}[!htbp]
		\centering
		\caption{An ablation study of our method in Experiment 3.}
		\begin{adjustbox}{width=6.5cm, height=1cm}
		\begin{tabular}{lll}
			\toprule
			\textbf{Methods} & \textbf{RMSE} & \textbf{MAE} \\
			\midrule
			ResNet-34 (CE) & $6.153\pm 1.071$ & $2.575\pm 0.477$ \\
			DA & $2.335\pm 0.269$ & $0.792\pm 0.078$ \\
			Ensemble & $1.983\pm 0.454$ & $0.843\pm 0.188$ \\
			DA+Ensemble  & $\bm{1.664\pm 0.275}$ & $\bm{0.621\pm 0.083}$ \\
			\bottomrule
		\end{tabular}%
		\end{adjustbox}
		\label{tab:EXP3_ablation_study}%
	\end{table}%

	\subsection{Experiment 4: Reduced Training Set}\label{sec:exp4}
	It is known that the performance of classification deteriorates when the number of training examples decreases. Therefore, this experiment is designed to test the effectiveness of the ensemble scheme when there are few training images for each distinct cell count, so the data augmentation is not used. In this experiment, we randomly delete half of the training images instead of removing all images for a certain cell count. Compared to Experiment 1, there are fewer training images for each distinct cell count under this setup. The whole experiment is repeated three times and the average performance of the candidate methods is shown in Table \ref{tab:EXP4_main_result}.	
	
	From Table \ref{tab:EXP2_ablation_study}, Table \ref{tab:EXP3_ablation_study}, and Table \ref{tab:EXP4_main_result}, we can see the ensemble scheme are able to improve the performance of ResNet-34 (CE) on both seen and unseen cell counts.  
	
	\begin{table}[!htbp]
		\centering
		\caption{Average test RMSE and MAE over three rounds in Experiment 4 with standard deviations after the ``$\pm$" symbol. The DA module in our framework is disabled because there is no missing cell counts.}
		\begin{tabular}{lll}
			\toprule
			\textbf{Methods} & \textbf{RMSE} & \textbf{MAE} \\
			\midrule
			DRDCNN \cite{liu2019novel} & $16.139\pm 22.767$ & $13.669\pm 19.772$ \\
			FPNCNN \cite{hernandez2018using} & $3.342\pm 0.516$ & $2.556\pm 0.381$ \\
			ERDCNN \cite{liu2019automated} & $3.531\pm 0.440$ & $2.715\pm 0.374$ \\
			ResNet-34 (MSE) \cite{xue2016cell} & $2.863 \pm 0.614$ & $2.124\pm 0.458$ \\
			\textbf{Ours} (only ResNet-34 (CE)) & $2.753 \pm 0.274$ & $1.017\pm 0.224$ \\
			\textbf{Ours} (Ensemble) & $\bm{1.969\pm 0.348}$ & $\bm{0.868\pm 0.196}$ \\
			\bottomrule
		\end{tabular}%
		\label{tab:EXP4_main_result}%
	\end{table}%

	\section{Conclusion}\label{sec:conclusion}
	
	In this work, we propose a novel framework to count cells from greyscale microscopic images without using manual annotations. We first formulate the cell counting problem as one of image classification and use classification-oriented ResNets to predict cell counts from images. Then, to deal with the two limitations of this formulation, we introduce simple but effective data augmentation and ensemble schemes. Our proposed framework achieves the state of the art on the Tiny-BBBC005 dataset and won the Case Study 1 competition of 47th Annual Meeting of the SSC.

	
	
	\bibliographystyle{IEEEtran}
	\bibliography{CellCountingBib}

	\clearpage
	\newpage
	\appendix
	
	\section*{Supplementary Material}
	\addcontentsline{toc}{section}{Supplementary Material}
	\renewcommand{\thesubsection}{S.\Roman{subsection}}
	\renewcommand\thetable{\thesubsection.\arabic{table}}
	

	%
	\subsection{Formulae for DA in Experiment 2 and 3}
	\label{supp:formulae}
	To test the performance of our proposed data augmentation method, we design experiment 2 and experiment 3 so that there are missing cell counts in the training set. In these two experiments, the proposed data augmentation method is applied to create synthetic training images for missing cell counts.  
	When conducting the data augmentation, for each missing count and each combination of stain type and blur level, we create $20$ synthetic images. Each synthetic image is created based on a randomly selected formula from a pool of $3-5$ formulae for each missing count.
	The formulae used in data augmentation in experiment 2 and experiment 3 are shown in Table~\ref{tab:EXP2_formulae} and Table~\ref{tab:EXP3_formulae} respectively. 
	
	\begin{table*}[!htbp]
		\footnotesize
		\centering
		\caption{Formulae for Each Deleted Cell Count in Experiment 2}
		\begin{tabular}{p{0.5cm}p{2.5cm}p{2.5cm}p{2.5cm}p{2.5cm}p{2.5cm}}
			\toprule
			Cell Count & Formula 1 & Formula 2 & Formula 3 & Formula 4 & Formula 5 \\
			\midrule
			& \multicolumn{5}{c}{Round 1}   \\
			\cline{2-6}
			14    & 5$\times$(2)+1$\times$(4) & 10$\times$(1)+1$\times$(4) & 1$\times$(14) &       &  \\
			35    & 23$\times$(1)+10$\times$(1)+ 1$\times$(2) & 10$\times$(3)+5$\times$(1) & 31$\times$(1)+1$\times$(4) & 27$\times$(1)+5$\times$(1)+ 1$\times$(3) & 10$\times$(2)+5$\times$(3) \\
			57    & 27$\times$(1)+10$\times$(3) & 48$\times$(1)+5$\times$(1)+ 1$\times$(4) & 23$\times$(2)+10$\times$(1)+ 1$\times$(1) & 27$\times$(1)+10$\times$(1)+ 5$\times$(4) & 18$\times$(2)+10$\times$(2)+ 1$\times$(1) \\
			66    & 31$\times$(2)+1$\times$(4) & 23$\times$(2)+10$\times$(2) & 61$\times$(1)+5$\times$(1) & 53$\times$(1)+10$\times$(1)+ 1$\times$(3) & 44$\times$(1)+10$\times$(1)+ 5$\times$(2)+1$\times$(2) \\
			83    & 40$\times$(1)+23$\times$(1)+ 10$\times$(2) & 53$\times$(1)+27$\times$(1)+ 1$\times$(3) & 78$\times$(1)+5$\times$(1) & 48$\times$(1)+31$\times$(1)+ 1$\times$(4) & 40$\times$(2)+1$\times$(3) \\
			\midrule
			& \multicolumn{5}{c}{Round 2}\\
			\cline{2-6}
			10    & 5$\times$(1)+1$\times$(5) & 5$\times$(2) & 1$\times$(10) &       &  \\
			31    & 14$\times$(2)+1$\times$(3) & 27$\times$(1)+1$\times$(4) & 18$\times$(1)+5$\times$(2)+ 1$\times$(3) & 23$\times$(1)+5$\times$(1)+ 1$\times$(3) &  \\
			70    & 40$\times$(1)+27$\times$(1)+ 1$\times$(3) & 66$\times$(1)+1$\times$(4) & 35$\times$(2) & 48$\times$(1)+18$\times$(1)+ 1$\times$(4) & 44$\times$(1)+18$\times$(1)+ 5$\times$(1)+1$\times$(3) \\
			83    & 44$\times$(1)+35$\times$(1)+ 1$\times$(4) & 78$\times$(1)+5$\times$(1) & 40$\times$(2)+1$\times$(3) & 66$\times$(1)+5$\times$(3)+ 1$\times$(2) & 35$\times$(1)+27$\times$(1)+ 5$\times$(4)+1$\times$(1) \\
			91    & 40$\times$(1)+27$\times$(1)+ 23$\times$(1)+1$\times$(1) & 87$\times$(1)+1$\times$(4) & 44$\times$(1)+40$\times$(1)+ 5$\times$(1)+1$\times$(2) & 74$\times$(1)+14$\times$(1)+ 1$\times$(3) & 78$\times$(1)+5$\times$(2)+ 1$\times$(3) \\
			\midrule
			& \multicolumn{5}{c}{Round 3}\\
			\cline{2-6}
			18    & 14$\times$(1)+1$\times$(4) & 10$\times$(1)+5$\times$(1)+ 1$\times$(3) & 10$\times$(1)+1$\times$(8) &       &  \\
			27    & 14$\times$(1)+10$\times$(1)+ 1$\times$(3) & 10$\times$(2)+5$\times$(1)+ 1$\times$(2) & 23$\times$(1)+1$\times$(4) &       &  \\
			44    & 23$\times$(1)+14$\times$(1)+ 5$\times$(1)+1$\times$(2) & 40$\times$(1)+1$\times$(4) & 14$\times$(1)+10$\times$(2)+ 5$\times$(2) & 35$\times$(1)+5$\times$(1)+ 1$\times$(4) & 31$\times$(1)+5$\times$(2)+ 1$\times$(3) \\
			53    & 40$\times$(1)+10$\times$(1)+ 1$\times$(3) & 48$\times$(1)+5$\times$(1) & 23$\times$(1)+10$\times$(3) & 48$\times$(1)+1$\times$(5) & 40$\times$(1)+10$\times$(1)+ 1$\times$(3) \\
			91    & 48$\times$(1)+40$\times$(1)+ 1$\times$(3) & 87$\times$(1)+1$\times$(4) & 70$\times$(1)+5$\times$(4)+ 1$\times$(1) & 78$\times$(1)+10$\times$(1)+ 1$\times$(3) & 61$\times$(1)+10$\times$(1)+ 5$\times$(4) \\
			\bottomrule
		\end{tabular}%
		\label{tab:EXP2_formulae}%
	\end{table*}%

	\begin{table*}[!htbp]
		\footnotesize
		\centering
		\caption{Formulae for Each Deleted Cell Count in Experiment 3}
		\begin{tabular}{p{0.5cm}p{2.5cm}p{2.5cm}p{2.5cm}p{2.5cm}p{2.5cm}}
			\toprule
			Cell Count & Formula 1 & Formula 2 & Formula 3 & Formula 4 & Formula 5 \\
			\midrule
			& \multicolumn{5}{c}{Round 1} \\
			\cline{2-6}
			61    & 35$\times$(1)+23$\times$(1)+ 1$\times$(3) & 31$\times$(1)+10$\times$(3) & 48$\times$(1)+10$\times$(1)+ 1$\times$(3) & 27$\times$(1)+23$\times$(1)+ 10$\times$(1)+1$\times$(1) & 40$\times$(1)+5$\times$(4)+ 1$\times$(1) \\
			66    & 27$\times$(1)+23$\times$(1)+ 10$\times$(1) +5$\times$(1)+1$\times$(1) & 48$\times$(1)+18$\times$(1) & 18$\times$(2)+10$\times$(2)+ 5$\times$(2) & 57$\times$(1)+5$\times$(1)+ 1$\times$(4) & 44$\times$(1)+10$\times$(1)+ 5$\times$(2)+1$\times$(2) \\
			70    & 40$\times$(1)+10$\times$(2)+ 5$\times$(2) & 35$\times$(2) & 40$\times$(1)+27$\times$(1)+ 1$\times$(3) & 44$\times$(1)+18$\times$(1)+ 5$\times$(1)+1$\times$(3) & 57$\times$(1)+10$\times$(1)+ 1$\times$(3) \\
			74    & 27$\times$(1)+23$\times$(1)+ 14$\times$(1)+10$\times$(1) & 44$\times$(1)+27$\times$(1) +1$\times$(3) & 57$\times$(1)+14$\times$(1)+ 1$\times$(3) & 53$\times$(1)+18$\times$(1)+ 1$\times$(3) & 48$\times$(1)+14$\times$(1)+ 10$\times$(1)+1$\times$(2) \\
			78    & 40$\times$(1)+35$\times$(1)+ 1$\times$(3) & 57$\times$(1)+18$\times$(1)+ 1$\times$(3) & 27$\times$(2)+14$\times$(1)+ 10$\times$(1) & 53$\times$(1)+23$\times$(1)+ 1$\times$(2) & 48$\times$(1)+14$\times$(1)+ 10$\times$(1)+5$\times$(1)+1$\times$(1) \\
			\midrule
			& \multicolumn{5}{c}{Round 2} \\
			\cline{2-6}
			70    & 40$\times$(1)+10$\times$(2)+ 5$\times$(2) & 35$\times$(2) & 40$\times$(1)+27$\times$(1)+ 1$\times$(3) & 44$\times$(1)+18$\times$(1)+ 5$\times$(1)+1$\times$(3) & 57$\times$(1)+10$\times$(1)+ 1$\times$(3) \\
			74    & 27$\times$(1)+23$\times$(1)+ 14$\times$(1)+10$\times$(1) & 44$\times$(1)+27$\times$(1)+ 1$\times$(3) & 57$\times$(1)+14$\times$(1)+ 1$\times$(3) & 53$\times$(1)+18$\times$(1)+ 1$\times$(3) & 48$\times$(1)+14$\times$(1)+ 10$\times$(1)+1$\times$(2) \\
			78    & 40$\times$(1)+35$\times$(1)+ 1$\times$(3) & 57$\times$(1)+18$\times$(1)+ 1$\times$(3) & 27$\times$(2)+14$\times$(1)+ 10$\times$(1) & 53$\times$(1)+23$\times$(1)+ 1$\times$(2) & 48$\times$(1)+14$\times$(1)+ 10$\times$(1)+5$\times$(1)+1$\times$(1) \\
			83    & 27$\times$(1)+23$\times$(2)+ 10$\times$(1) & 53$\times$(1)+10$\times$(2)+ 5$\times$(2) & 27$\times$(1)+23$\times$(1)+ 18$\times$(1)+10$\times$(1)+ 5$\times$(1) & 66$\times$(1)+10$\times$(1)+ 5$\times$(1)+1$\times$(2) & 40$\times$(2)+1$\times$(3) \\
			87    & 44$\times$(1)+23$\times$(1)+ 10$\times$(2) & 57$\times$(1)+14$\times$(1)+ 10$\times$(1)+5$\times$(1)+ 1$\times$(1) & 66$\times$(1)+10$\times$(1)+ 5$\times$(2)+1$\times$(1) & 44$\times$(1)+23$\times$(1)+ 10$\times$(1)+5$\times$(2) & 48$\times$(1)+10$\times$(2)+ 5$\times$(3)+1$\times$(4) \\
			\midrule
			& \multicolumn{5}{c}{Round 3} \\
			\cline{2-6}
			83    & 27$\times$(1)+23$\times$(2)+ 10$\times$(1) & 53$\times$(1)+10$\times$(2)+ 5$\times$(2) & 27$\times$(1)+23$\times$(1)+ 18$\times$(1)+10$\times$(1)+ 5$\times$(1) & 66$\times$(1)+10$\times$(1)+ 5$\times$(1)+1$\times$(2) & 40$\times$(2)+1$\times$(3) \\
			87    & 44$\times$(1)+23$\times$(1)+ 10$\times$(2) & 57$\times$(1)+14$\times$(1)+ 10$\times$(1)+5$\times$(1)+ 1$\times$(1) & 66$\times$(1)+10$\times$(1)+ 5$\times$(2)+1$\times$(1) & 44$\times$(1)+23$\times$(1)+ 10$\times$(1)+5$\times$(2) & 48$\times$(1)+10$\times$(2)+ 5$\times$(3)+1$\times$(4) \\
			91    & 61$\times$(1)+27$\times$(1)+ 1$\times$(3) & 57$\times$(1)+14$\times$(1)+ 10$\times$(2) & 27$\times$(2)+23$\times$(1)+ 14$\times$(1) & 78$\times$(1)+10$\times$(1)+ 1$\times$(3) & 61$\times$(1)+10$\times$(1)+ 5$\times$(4) \\
			96    & 40$\times$(1)+27$\times$(1)+ 23$\times$(1)+5$\times$(1)+ 1$\times$(1) & 74$\times$(1)+10$\times$(2)+ 1$\times$(2) & 35$\times$(1)+31$\times$(1)+ 5$\times$(6) & 78$\times$(1)+18$\times$(1) & 74$\times$(1)+14$\times$(1)+ 5$\times$(1)+1$\times$(3) \\
			100   & 27$\times$(2)+23$\times$(2) & 70$\times$(1)+10$\times$(2)+ 5$\times$(2) & 44$\times$(1)+18$\times$(2)+ 10$\times$(2) & 78$\times$(1)+18$\times$(1)+ 1$\times$(4) & 74$\times$(1)+14$\times$(1)+ 5$\times$(2)+1$\times$(2) \\
			\bottomrule
		\end{tabular}%
		\label{tab:EXP3_formulae}%
	\end{table*}%
	
\end{document}